\documentclass[final]{agujournal2019}
\makeatletter
\AtBeginDocument{
  \def\@oddhead{\vbox to 0pt{\vss\centerline{}\vskip24pt}}
  \let\@evenhead\@oddhead
}
\makeatother

\usepackage{url} 
\usepackage{soul}
\linenumbers
\draftfalse
\journalname{Geophysical Research Letters}

\nolinenumbers

\begin{document}
\title{Ocean surface gravity waves excited by the 2022 eruption of Hunga Tonga-Hunga Ha'apai Volcano}
\thanks{
Accepted manuscript of:\\
Nishida, K., Ichihara, M., Kubota, T., and Tonegawa, T. (2024),``Ocean Surface Gravity Waves Excited by the 2022 Eruption of Hunga Tonga-Hunga Ha'apai Volcano'', \textit{Geophysical Research Letters}, \textbf{51}, e2024GL111983.\\
Published version:
https://doi.org/10.1029/2024GL111983\\
This manuscript corresponds to the peer-reviewed accepted version of the article.
}
\authors{Kiwamu Nishida\affil{1}, Mie Ichihara\affil{1}, Tatsuya Kubota\affil{2}, Takashi Tonegawa\affil{3}}
\affiliation{1}{Earthquake Research Institute, the University of Tokyo}
\affiliation{2}{National Research Institute for Earth Science and Disaster Resilience, Tsukuba, Japan}
\affiliation{3}{Japan Agency for Marine-Earth Science and Technology (JAMSTEC), Yokosuka, Japan}
\correspondingauthor{Kiwamu Nishida}{knishida@eri.u-tokyo.ac.jp}

\begin{keypoints}
\item Ocean surface gravity waves at 15--40\,mHz were excited at 4:00 UTC when an underwater eruption occurred at Hunga Tonga-Hunga Ha'apai Volcano.
\item A significant sub-event at about 8:40 UTC followed the main event, and the ocean surface gravity waves lasted for half a day.
\item The observations of ocean surface gravity waves suggest a blowout of the seawater above the summit and the inflow of seawater into the crater.
\end{keypoints}

\begin{abstract}
 On 15 January 2022, a massive underwater eruption occurred at the Hunga Tonga-Hunga Ha'apai Volcano. The plume reached the mesosphere, and the eruption excited a significant atmospheric Lamb wave, which forced the tsunami. The complicated tsunami waveforms due to ocean-atmosphere coupling prevented inferring the force history of the excitation. To address this, we analyze ocean surface gravity waves (OSWs) from 15 to 40\,mHz, which are decoupled from the Lamb wave due to their slower phase velocities.  Modeling these OSWs, we infer that the excitation started at 4:00 UTC with an amplitude of $10^{10}$\,N and lasted for 5 hours, followed by a sub-event at 8:40 UTC. The observations suggest an initial blowout of seawater above the summit and a subsequent outflow that excited a tsunami below 5 mHz. The 2-hour delayed OSW excitation from 6 to 15\,mHz may indicate seawater inflow into the crater.
\end{abstract}

\section*{Plain Language Summary}
On 15 January 2022, the Hunga Tonga-Hunga Ha'apai volcano experienced a massive underwater eruption. This eruption sent a plume into the atmosphere and produced a powerful atmospheric wave that caused an early arrival of the tsunami. The complex tsunami propagation coupled with the atmospheric wave prevents us from understanding the excitation processes. To address this, we focussed on ocean surface gravity waves (OSWs) in an oscillatory period from 25 to 67\,s, also interpreted as short-period tsunamis. These OSWs are less affected by these atmospheric waves because of the slow propagation speed. Because the OSWs are less influenced by ocean depth, they are easier to model. The observed OSWs show that the eruption started at 4:00 UTC with a vertical force of approximately $10^{10}$\,N and continued for about 5 hours. Another significant sub-event occurred around 8:40 UTC. Furthermore, the OSWs in an oscillatory period from 67 to 167\,s arrived about two hours after the short-period OSWs. These observations suggest that the eruption blew seawater above the volcano’s summit and a subsequent outflow that excited a tsunami. The 2-hour delay in low-frequency OSW could indicate that seawater flowed into the crater.

\section{Introduction}

The Hunga Tonga-Hunga Ha'apai volcano (hereafter noted as ``HTHH'') erupted on 15 January 2022. The underwater eruption was the largest in the last 100 years. It generated various waves in the solid Earth, the ocean, and the atmosphere, with amplitudes large enough to be recorded in modern geophysical observations, which was previously inconceivable. The coupling of the atmosphere, ocean, and solid earth played a particularly important role in this eruption. The eruption excited huge atmospheric Lamb waves, which orbited Earth many times \cite<e.g.,>{Matoza2022-tt,Omira2022-al,Yuen2022-na}. The amplitude was so huge that the thermal anomalies associated with the Lamb wave propagation were captured by satellite images \cite{Wright2022-lh} and surface meteorological stations \cite{Watada2023-eb}. The Lamb waves forced tsunamis with arrivals earlier than normal tsunamis, which were observed around the world \cite<e.g.,>{Kubota2022-fb,Omira2022-al,Nishikawa2022-ep,Mizutani2023-gb,Tonegawa2023-gm}. Pekeris mode, a theoretically predicted atmospheric boundary wave, was also detected for the first time \cite{Watanabe2022-bu}. Ionospheric total electron content (TEC) from Global Navigation Satellite System (GNSS) observation showed acoustic wave propagation in the ionosphere \cite<e.g.,>{Astafyeva2022-xy,Ravanelli2023-xf}. The eruption also excited acoustic coupling modes between the solid earth and the atmosphere at 3.7 and 4.4\,mHz, which lasted several hours \cite{Matoza2022-tt, Ringler2022-ti,Garza-Giron2023-bh}. 

We focus on the time history of the excitation force (hereafter, source-time function) for the tsunami to understand the eruption processes. We can infer it if we know the exact Green function. However, the early arrival of the observed tsunamis can be attributed to the forcing by atmospheric Lamb waves. The coupling with atmospheric Lamb waves complicated the ocean wave propagation below 1\,mHz. The tsunami also lasted for days, which can be explained by the energy transfer from atmospheric waves through the coupling \cite{Kubo2022-le}. The strong atmosphere-ocean coupling made it difficult to infer the excitation force quantitatively.

A tsunami is a kind of ocean surface gravity wave (OSW). In general, OSW follows the dispersion relation as
\begin{equation}
    c_p = \sqrt{gd \frac{\tanh kd}{kd}},
\end{equation}
where $c_p$ is the phase velocity, $g$ is gravitational acceleration, $k$ is the wavenumber, and $d$ is the water depth. A tsunami is also often treated as a shallow-water wave because its wavelength is typically longer than the water depth. The phase velocity of the shallow-water wave can be approximated by $\sqrt{gd}$ as the low-frequency limit. At higher frequencies with a wavelength shorter than the ocean depth (typically above 10\,mHz), OSW is treated as a deep-water wave. The phase velocity is proportional to the reciprocal frequency as $c_p = g/\omega$. Ocean swell is known to be an example of a deep-water wave. Dispersive OSWs are also excited during a distant large storm \cite<e.g.,>{Tonegawa2018-zp}. 

After the HTHH eruption, OSWs in the deep-water wave regime were also reported \cite{Le_Bras2023-gs, Diaz2023-cc}. The arrival times of the OSWs after the HTHH eruption were delayed by about four days for a distance of 10,000 km at 30\,mHz because the group velocity of the deep-water wave $c_g$ also decreases with frequency as $g/(2\omega)$. Because the slow propagation speed weakens the coupling to atmospheric waves, the propagation of the OSW above 10\,mHz becomes simple. 
In addition, OSWs in the deep-water wave regime have energy only near the sea surface (typically shallower than 1\,km), so that they are less sensitive to the ocean bottom boundary. Therefore, propagation modeling in the deep-water wave regime is simpler than for tsunamis in the shallow-water wave regime. The simple propagation properties enable us to infer the source-time function accurately. This study aims to reveal the eruption force system by simple modeling of the dispersive OSW in a deep-water wave regime decoupled from atmospheric waves. 

\begin{figure}[ht]
    \centering
    \includegraphics[width=15cm]{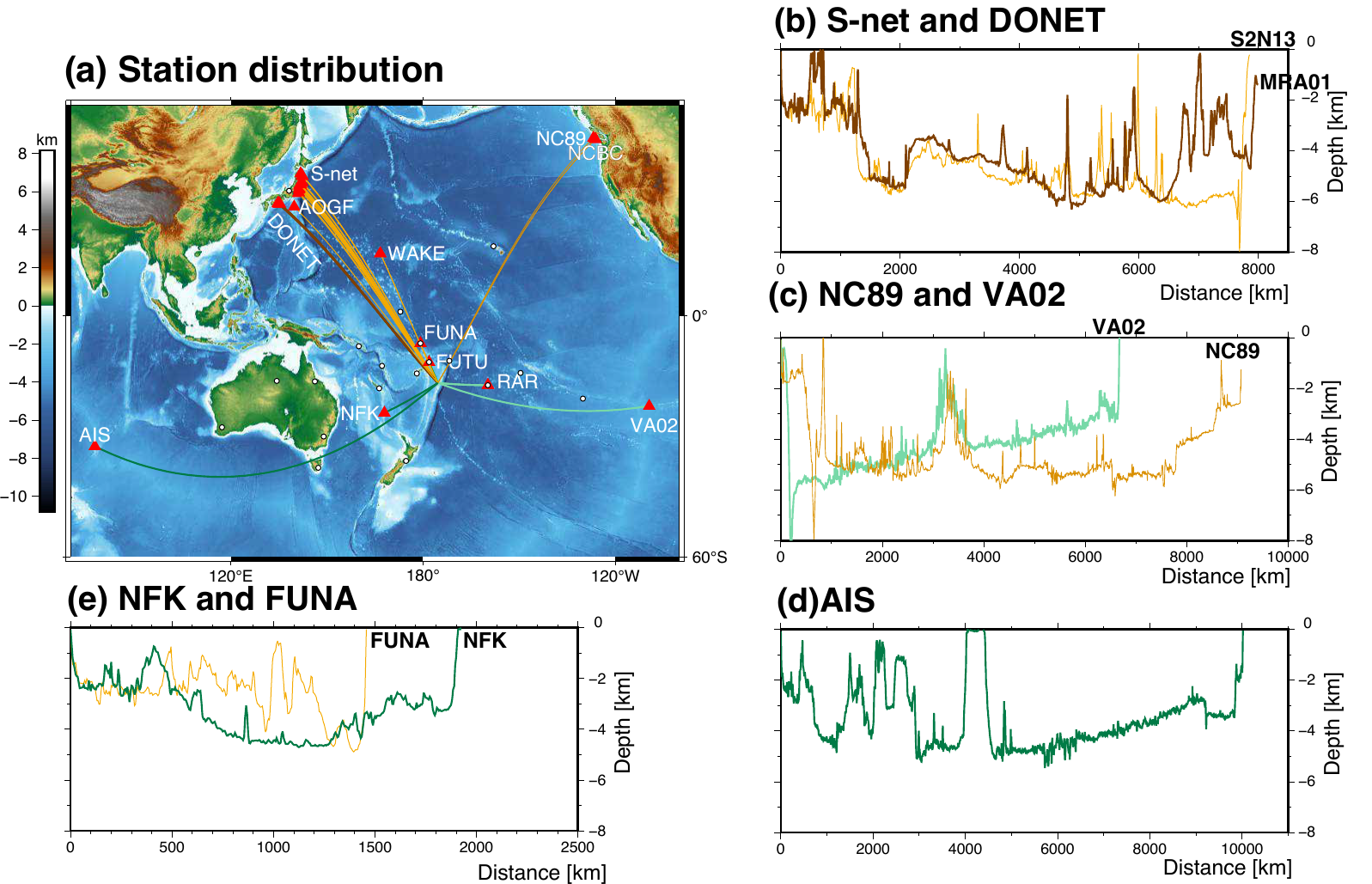}
    \caption{(a) Station distribution. Red triangles show stations that observed the OSWs with the names and the ray paths classified by the colors. Open circles show 19 broadband seismic stations (CAN, FUTU, KIP NOUC, PPTF, SANVU, MSVF, TAU, WRAB, AFI, CTAO, FUNA, HNR, MAJO, NWAO, PTCN, RAR, SNZO, TARA) used for the Rayleigh wave analysis. 
    (b--e) Ocean depth sections along the ray paths from the volcano to an S-net station (S2N13) and a DONET station (MRA01) in Japan (b); to VA02 (Isla de Pascua, Chile) and a NEPTUNE station NC89 in Canada (c);  to AIS (Nouvelle-Amsterdam, France) (d); and to NFK (Norfolk Island, Australia) and FUNA (Funafuti, Tuvalu) (e).
    }
    \label{station}
\end{figure}

\section{Observations}
We analyze data from ocean bottom pressure gauges in Japan (S-net and DONET) and Canada (NEPTUNE) shown in Figure \ref{station}a. Figures \ref{station}b--e show bathymetry along the ray paths, and the water depth is deeper than 2\,km in most parts. We also analyzed data from island broadband seismic stations because they sometimes record seismic deformation associated with OSWs \cite<e.g., the tsunami after the 2004 Sumatra-Andaman earthquake>{Yuan2005-ir,Okal2007-sc,Okal2007-vn} and the 2010 Mentawai tsunami earthquake \cite{Nishida2019-eb}.

\begin{figure}[h]
    \centering
    \includegraphics[width=15cm]{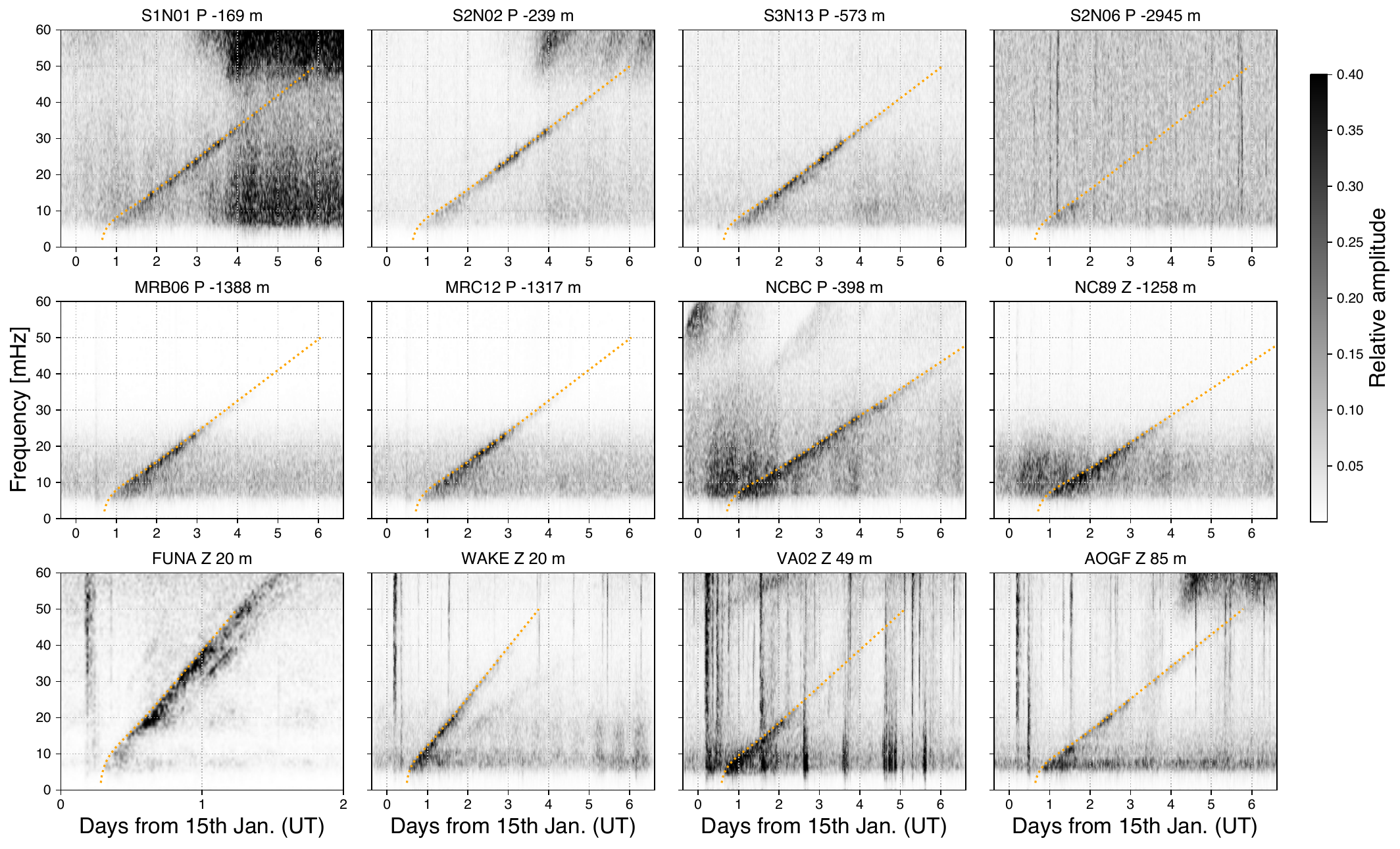}
    \caption{
    Running spectra at all the stations. ``P'' represents ocean bottom pressure at DONET, S-net, and NEPTUNE, respectively, whereas ``Z'' represents vertical ground velocity recorded by broadband seismometers. We also show the station elevation in the titles. The horizontal axes show days from 15 January 2022. The theoretical travel times are also shown by the orange dashed lines. S1N01, S2N02, S3N13 and S2N06 are S-net stations, MRB06 and MRC12 are DONET stations. Other stations are shown in the Figure \ref{station}a.
    }
    \label{spectrogram1}
\end{figure}

Figure \ref{spectrogram1} shows typical examples of spectrograms at the stations. We calculated the spectrograms of the ocean bottom pressure gauges and the vertical components of the broadband seismometers. Each spectrogram shows a slanted line of arrival times as a function of the frequency, which can be explained by the group velocity $g/(2\omega)$ of the deep-water wave above about 10\,mHz \cite{Le_Bras2023-gs,Hanson2005-cz}. Below 5\,mHz, the arrival time converges to that of the shallow water wave (i.e., the arrival of the tsunami).
The high-frequency component gradually disappears with increasing depth of pressure gauges (MRB06, MRC12, S2N06) beyond 1000\,m (Figure \ref{spectrogram1}).
In contrast, the stations at depths of about 100--200\,m (S1N01, S2N02)  show persistent high energy above 40\,mHz. The onlnad stations also show the high-frequency energy, which can be explained by direct loads of ocean swell activities \cite<e.g.,>{Janiszewski2022-zp,Webb2010-vb}. All panels show no significant later phases of OSWs. As explained in the next section, the simple propagation property enables us to synthesize it with a simple model. 

\section{Estimation of the source-time function}
We estimated the source-time function by modeling the observed ocean bottom pressure of OSW. Based on previous studies on source inversion using teleseismic waves \cite{Garza-Giron2023-bh,Thurin2023-gw,Poli2022-ef}, we assumed that the excitation processes could be approximated by a vertical single-force at the centroid location, which can be interpreted as the momentum flux of the erupted mass and the pressure of the eruptive jet \cite<e.g.,>{Brodsky1999-hw}.

The source-time function $\phi^{OSW}_{p,i}(t)$ at time $t$ is estimated by modeling the ocean bottom pressure recorded by the $i$th station of DONET and S-net. We deconvolve the source-time function the observed pressure $P_i(\omega)$ in the frequency domain with the Green function $G^{OSW}_{p,i}(\omega)$, where $\omega$ is the angular frequency. The source-time function is given by the inverse Fourier transform as
\begin{equation}
    \phi^{OSW}_{p,i}(t) = \frac{1}{2\pi}\int_{\omega_{min}}^{\omega_{max}} \frac{P_i (\omega)}{G^{OSW}_{p,i}(\omega)} e^{\mathrm{i}\omega t} d\omega,
\end{equation}
where $\omega$ is angular frequency, $\omega_{min}$ corresponds to 15\,mHz and $\omega_{max}$ corresponds to 40\,mHz.
Although this frequency range is near the deep-water wave regime, travel time calculations with an accuracy of a few minutes require considering bathymetric changes along the ray path. The Green function $G^{OSW}_{p,i}$ for a vertical single-force is calculated based on ray theory with an assumption that the ocean depth changes slowly compared to the wavelength \cite<WJKB approximation>[see Appendix A for details]{Dahlen1998-hr}.
For simplicity, the ray path is also assumed to be straight. We have validated the straight-ray assumption by ray tracing using an Eikonal solver \cite{White2020-oc}. The phase travel-time anomalies of the selected S-net stations are within 5\,min from 15 to 40\,mHz, whereas those of DONET are within 30\,min from 15 to 40\,mHz. The larger delays of the DONET stations can be explained by ray bending because the DONET ray paths experience shallower depths in Fiji islands near the source, as shown in Figure \ref{station}b. This approximation allows for minute-precision travel time calculations for S-net stations, but it is still difficult to match the phases for different stations. Therefore, this study will consider envelopes of the source-time function \cite<e.g.,>{Vergoz2022-ch}.

\begin{figure}
    \centering
    \includegraphics[width=12cm]{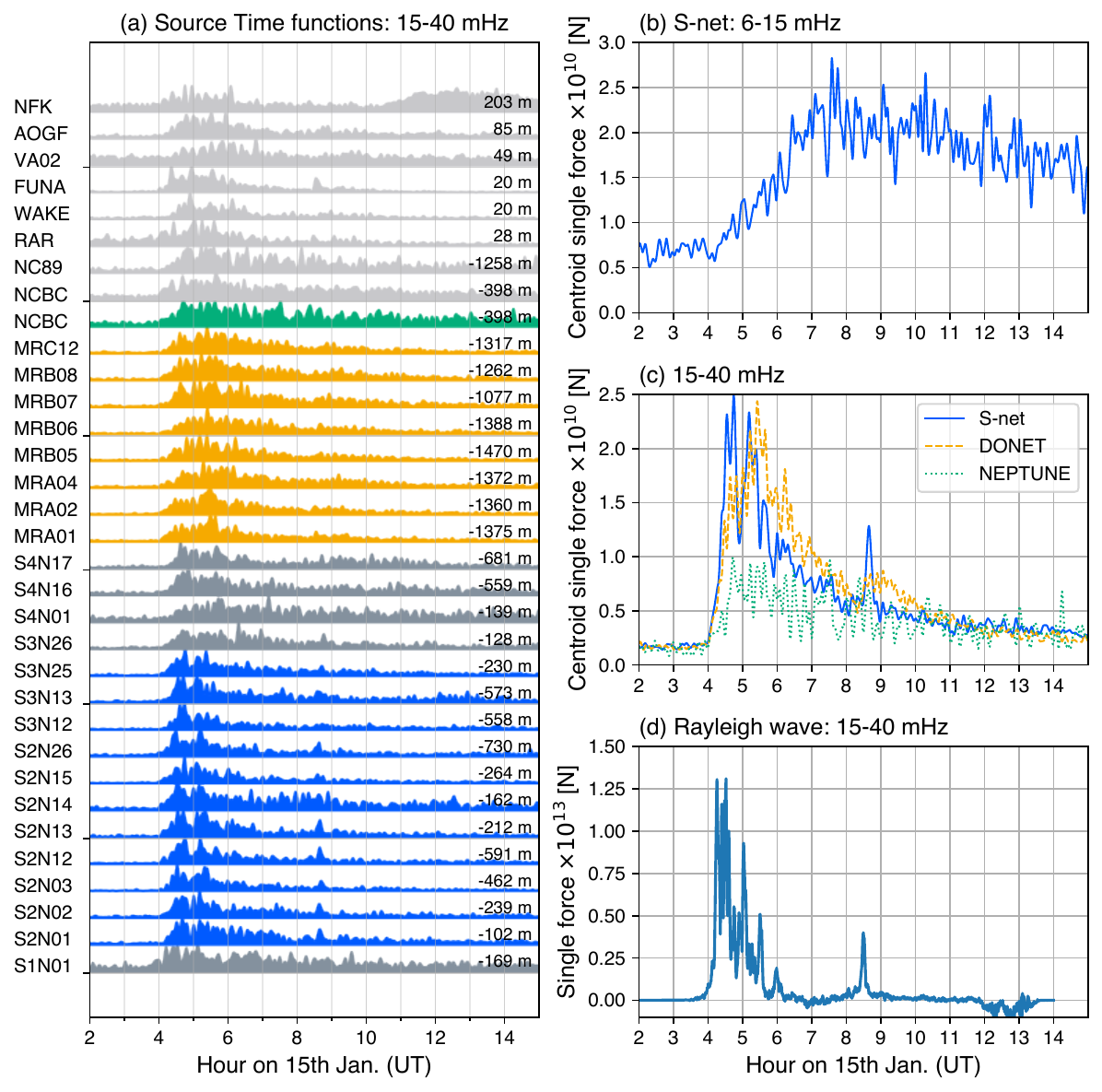}
    \caption{(a) Estimated source-time function of OSWs from 15 to 40\,mHz for each station. The curves from the lower NCBC to S1N01 (green, orange, dark gray, and blue) show the estimated ones using ocean bottom pressure gauges, whereas the light gray ones above the upper NCBC show the estimation using island broadband seismometers by back-projecetion to the source. 
   The envelopes were low-pass filtered at 2\,mHz. The amplitudes are normalized by the maximum value. 
   We also show the station elevation on the right.
   (b) Estimated source-time functions of OSWs stacked in  6--15\,mHz over S-net. (c) Estimated source-time functions of OSWs in 15--40\,mHz stacked over S-net (blue solid line) and DONET (orange dashed line). We also plot the envelope of a NEPTUNE station NCBC (green dashed line). The stations used are shown in the same color in Figure \ref{STF}a. (d) Estimated source-time function of Rayleigh waves in 15--40\,mHz using 19 on-land broadband stations shown in Figure \ref{station}a. 
    }
    \label{STF}
\end{figure}

Figure \ref{STF}a shows the envelopes of the source-time function $\phi^{OSW}_{p,i}(t)$. For reference, the figure also shows envelopes of backprojections of island broadband seismic data (indicated with ``Z'' in Figure \ref{spectrogram1}) from the stations to the source, whose profiles are consistent with $\phi^{OSW}_{p,i}(t)$. However, amplitudes are difficult to infer because the seismic deformation of the island due to the loads by the OSWs was too complicated to model at high frequency \cite{Nishida2019-eb}.

All the envelopes show the onset at around 4:00 UTC and the two maxima before and after 5:00 UTC. At around 8:40 UTC, the S-net stations show a significant sub-event. As discussed later, the sub-event was dominated at a higher frequency above 20\,mHz. DONET results lack the sub-event because the greater depth at the DONET stations attenuated the high-frequency component of the sub-event. A broadband seismic station at FUNA (Funafuti, Tuvalu) with an arc distance of 13 degrees, also shows the sub-event. The onset times of both the main eruption and the sub-event were consistent with those of the S-net stations (about 70 degrees). The consistency suggests that the back-projection for the source-time function is appropriate even at the distant stations.

Because the estimated single-force at each station has a larger error, we stacked the envelopes with a low-pass filter of 2\,mHz over the S-net and DONET stations, respectively, in Figure \ref{STF}c, which also shows the envelope of NCBC (Canada) for comparison. Although the noise level of the unstacked data at station NCBC is larger, all the amplitudes are consistent with each other. Even after stacking, DONET lacks the sub-event due to the greater depth of the stations. Because the Fiji islands are located between the great circle path between the source and DONET stations, scattering in shallow depth caused codas, which blurred the source-time function. Because the ray paths of the 5 S-net stations (S1N01, S3N26, S4N01, S4N16, S4N17) also intersect an island in the middle, the larger codas affect the estimated source-time functions. For further discussion, we carefully chose source-receiver pairs (blue color in Figure \ref{STF}a) where the ray paths did not intersect islands. The estimated source-time functions show a consistent peak time with the ambiguity of the modeling of about 5\,min. We will discuss amplitudes of the vertical single-force based on the most reliable S-net results.
The forcing of OSW (blue color in Figure \ref{STF}c) started at 4:00 UTC and reached a maximum of about $2\times 10^{10}$\,N at around 4:45 UTC. The forcing gradually attenuated with some sub-peaks, but it lasted for half a day. At 8:40 UTC, the source-time function shows an isolated sub-event with an amplitude of about $1\times 10^{10}$\,N.
 
To compare the excitation of Rayleigh waves, we also calculated the source-time function of Rayleigh waves. 
By deconvolving the corresponding Green's function, we estimated the source-time function $\phi^{Ray}_{r,i}(t)$ and $\phi^{Ray}_{z,i}(t)$ for the vertical single-force from the radial and vertical components of $i$th station, respectively.
The stacking of $\phi^{Ray}_{r,i}(t)\phi^{Ray}_{z,i}(t)$ over all seismic stations was bandpass-filter from 15 to 40\,mHz.
This procedure emphasizes the elliptic particle motions of the Rayleigh wave, because $\phi^{Ray}_{r,i}(t)$ and $\phi^{Ray}_{z,i}(t)$ should be identical. On the other hand, it suppresses the contribution of the body wave and the noise because $\phi^{Ray}_{r,i}(t)\phi^{Ray}_{z,i}(t)$ takes positive and negative values due to the phase difference.  
Figure \ref{STF}d shows the square root of the absolute value with the sign as the source-time function of seismic Rayleigh waves with a maximum amplitude of about $10^{13}$\,N, which is consistent with past studies \cite<e.g.,>{Garza-Giron2023-bh}. The seismic force is 3 orders of magnitude larger than that of the OSW.  The forcing abruptly attenuated after 6:00 UTC, whereas the forcing of OSW lasted for half a day. The difference suggests information on the explosion processes, as discussed later. 

The peak time of the source-time functions of the Rayleigh wave was slightly earlier than the peak time of the OSW. The onset time of the seismic sub-event at around 8:30 UTC was also slightly earlier than the corresponding sub-event of OSW (Figure \ref{STF}c and \ref{STF}d). The onset time of the seismic event is consistent with past studies \cite<e.g.,>{Vergoz2022-ch}. 
Because the seismic waves were also excited by the brittle fracture prior to the main eruption \cite{Horiuchi2024-fl}, the OSW excitation could be delayed. 

\begin{figure}[h]
    \centering
    \includegraphics[width=12cm]{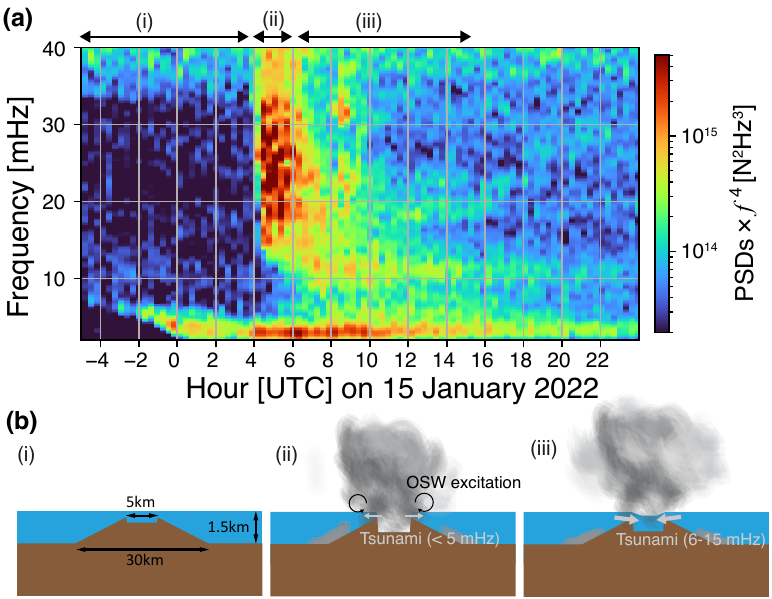}
    \caption{(a) Stacked spectrogram of source-time function estimated from S-net data. At each frequency, the PSDs were multiplied by $f^4$ to compare tsunami and OSW amplitudes. The large sub-event occurred at 8:40 UTC. 
    (b) A schematic figure showing the eruption sequence. (i) Before the eruption. (ii) The main eruption stage from 4:00--6:00 UTC. The caldera subsided up to 800\,m  \cite{Seabrook2023-nl}. (iii) The caldera was filled with water. 
    }
    \label{spectrogram2}
\end{figure}

Figure \ref{spectrogram2} shows a stacked running spectrum over the selected S-net stations. To compare the OSWs with tsunamis, we plot it from 2\,mHz. The main event above 15\,mHz occurred at 4:00 UTC, and the sub-event above 22\,mHz occurred at 8:40 UTC. 
Figure \ref{spectrogram2}a also shows that the source-time function of the sub-event around 8:40 UTC has resonant peaks at around 10, 20, and 30\,mHz. The resonant peaks suggest modal oscillations of trapped OSWs around the volcano \cite<e.g.,>{Longuet-Higgins1967-kr}. 
The excited time of the OSW above 15\,mHz was about 4:00 UTC. On the other hand, OSW from 6 to 15\,mHz was delayed for about two hours (Figure \ref{spectrogram2}a). Figure \ref{STF}b shows the source-time function from 6 to 15\,mHz stacked over the S-net data, which also shows the significant delay in the peak time. 

The tsunami below 5\,mHz is estimated to be excited before 0:00 UTC, that is before the 15 January eruption onset. However, the early arrival is an artifact because the tsunami was back-projected to the source assuming a dispersion of normal OSW (typically 200\,m/s in the deep ocean), which is significantly slower than the observed anomalous tsunami forced by non-dispersive atmospheric Lamb wave \cite{Kubota2022-fb}. Because the forced tsunami followed the Lamb wave propagations, the assumed dispersion of OSW apparently caused the arrival time earlier. 
After the apparent early arrival, the figure also shows a significant normal tsunami excited at 4:00 UTC \cite{Purkis2023-ec,Borrero2022-gk}. 
Although the tsunami at lower frequencies lasted for days \cite{Kubota2022-fb}, the duration of the OSWs above 10\,mHz is approximately half a day after removal of the dispersion effects. This observation also supports that the tsunami persistently received energy from atmospheric waves, not from the volcanic eruption \cite{Kubota2022-fb}.

\section{Discussions}
On the basis of the observations, let us discuss the physical processes of the eruption. The single-force of the seismic event is estimated to be on the order of $10^{13}$\,N, which is consistent with past studies \cite<e.g.,>{Garza-Giron2023-bh,Poli2022-ef}. 
They interpreted the seismic single-force to the ejected mass of $10^{13}$\,kg in $\sim$200\,s duration of the explosion, although there remains a great uncertainty about the assumptions. 
The ejected mass generating the single-force should partly contain seawater entrained during the eruption.
 The microwave Limb Sounder revealed water ejection of about $10^{11}$\,kg into the stratosphere \cite{Millan2022-xl,Schoeberl2022-tj}, which could be the lower limit of the seawater ejected into the atmosphere. Because the corresponding water volume is $10^8$\,m$^3$, we consider that the water above the summit was blown out at the primary seismogenic explosion (Figure \ref{spectrogram2}b). 

Only a small part of the ejection contributed to the excitation of the OSWs in the 15--40\,mHz range because the force to excite the OSWs is much smaller than the seismic force. 
If the eruption ejects seawater faster than OSWs can, and the overpressure of the high-speed expanding jet sustains the vertical seawater wall above the summit shown in (ii) of Figure \ref{spectrogram2}b, the OSW excitation could be small. 

We verify the estimated amplitude of the equivalent vertical single-force to excite the OSW using a simple model. The OSW is assumed to be excited by random atmospheric pressure $\delta p$ with correlation length $L$ around the crater. The pressure cells were distributed on the area $S$. The equivalent vertical single-force is estimated to be $\sqrt{N}F$, where $N$ is the number of the cells $S/\pi L^2$, and $F$ is force per one cell given by $\delta p \pi L^2$. 
For example, using $S=7.5\times 10^7$\,m$^2$, $\delta p=2\times 10^3$\,Pa, and $L=500$\,m, the estimated equivalent single-force of $1.5\times 10^{10}$\,N remains consistent with the observations, despite the uncertainty in these parameters. This stage might last for 2 hours during the large seismic excitation. 

Figure \ref{STF} also shows the excitation of the normal tsunami at 4:00 UTC. 
The single-force source makes it difficult to excite the observed tsunami below 5\,mHz, because the OSW excitation by the single-force is less efficient at low frequencies (see Appendix A).
The excitation may be explained by expanding the internal wall. If the expansion time scale of the internal wall is 200\,s, the tsunami excitation above 5\,mHz should be small.
This model can also explain the observed lack in the OSWs from 6 to 15\,mHz.

The lightning observation indicates that the eruption declined by 6:00 UTC or 6:30 UTC \cite{Ichihara2023-zr}. 
The 2-hour delayed excitation of OSWs may indicate the inflow of seawater into the corrupted crater. The spectrogram in Figure \ref{spectrogram2}a also shows that the source-time function of the sub-event around 8:40 UTC has resonant peaks at around 10, 20, and 30\,mHz. The resonant peaks suggest the trapping of OSW in the seawater above the summit \cite<e.g.,>{Longuet-Higgins1967-kr}. 

\section{Conclusions}
We analyze OSWs recorded by ocean bottom pressure gauges and island broadband stations.  The source-time function of the OSWs excited by the 2022 eruption of HTHH was estimated using the pressure gauge data. The OSWs were modeled by a convolution between the vertical single-force and the ray theoretical Green function. The forcing from 15 to 40\,mHz started at 4:00 UTC and reached a maximum of about $10^{10}$\,N at around 4:30 UTC. The forcing gradually attenuated, but it lasted for half a day. At 8:40 UTC, the source-time function also shows the largest sub-event. The estimated single-force is 3 orders of magnitude smaller than that of the seismic wave. Furthermore, tsunamis from 6 to 15\,mHz were also found to be excited about 2 hours after the OSW excitation. These observations suggest an initial blowout of seawater above the summit and a subsequent outflow that excited a tsunami below 5 mHz. The 2-hour delayed OSW excitation from 6 to 15\,mHz may indicate seawater inflow into the crater.

\appendix
\section{The Green function of OSW for a vertical single-force} 

The Green function for a vertical single-force can be represented by
\begin{equation}
   G^{OSW}_{p,i}(\omega)= 
   U(R_e,\hat{\mathbf{r}}_s,\omega) 
   U(R_{o,i},\hat{\mathbf{r}}_i,\omega) 
   \frac{e^{-\mathrm{i}\left(\int_0^{\Delta_i} k(\Delta,\omega)d\Delta -\frac{\pi}{4}\right)}}{\sqrt{8\pi k(\Delta_i,\omega) S_i R_e}}
\end{equation}
where  $k(\Delta,\omega)$  is the local wave number along the ray with the arc distance $\Delta$ from the source, $\Delta_i$ is the arc distance between the source and the $i$th station, $S_i$ is the geometrical spreading at $i$th station, $U$ is the eigenfunction of OSW in the vertical component with a polar coordinate, $R_{o,i}$ is the radius of the ocean bottom at the $i$th station, $R_e$ is the Earth’s radius, $\hat{\mathbf{r}}_s$ shows the source location on the Earth's surface $(\theta_s,\varphi_s)$ in a polar coordinate, and $\hat{\mathbf{r}}_i$ shows the $i$th station $(\theta_i,\varphi_i)$.
To calculate the local wavenumber, we used a 1 arc-minute relief model ETOPO1 \cite{Amante2009-jn}. Following \citeA{Tromp1992-qt} the eigenfunctions $U(r, \theta,\varphi,\omega)$ at a point $(r,\theta,\varphi)$ using the polar coordinate are normalized as $c (\hat{\mathbf{r}},\omega)C(\hat{\mathbf{r}},\omega) I_1(\hat{\mathbf{r}},\omega) = R_e^2$, where $c$ is the phase velocity in km/s, $C$ is the group velocity in km/s. The energy integral $I_1$  defined in \cite{Tromp1992-qt} can be approximated as
\begin{equation}
    I_1(\hat{\mathbf{r}}, \omega) 
    \approx  \frac{R_e^2}{2}\frac{\rho g U^2(R_e,\hat{\mathbf{r}}, \omega)}{\omega^2}, 
\end{equation}
because the kinetic energy of OSW per unit area is given by $\rho g U^2/4$ \cite<e.g.,>{Pedlosky2003-va}. $G^{OSW}_{p,i}$ is proportional to $\omega^3$ in a frequency range of deep-water wave if the station depth is shallower than the wavelength, whereas $G^{OSW}_{p,i}$ is proportional to $\omega^{3/2}$ in the frequency range of shallow-water wave. 

\section*{Open Research Section}
We downloaded seismic and pressure gauge data through EarthScope Consortium Web Services (\url{https://service.iris.edu/}), including the following seismic networks: (1) IU (Albuquerque Seismological Laboratory \url{10.7914/SN/IU});(2) AU (Australian National Seismograph Network Data Collection, \url{10.26186/1446750}); (3) C1 (Red Sismologica Nacional \url{10.7914/SN/C1}); (4) G (GEOSCPE, \url{10.18715/GEOSCOPE.G}); (5) NEPTUNE seismic stations (Ocean Networks Canada., 2009, \url{10.7914/SN/NV}). We used S-net, DONET, and F-net data obtained by the National Research Institute for Earth Science and Disaster Resilience. The S-net and DONET when the 2022 eruption of HTHH is also available at \url{doi.org/10.17598/NIED.0007-2022-001} and \url{doi.org/10.17598/NIED.0008-2022-001}.
Pre- and post-processing was performed using ObsPy \cite{Krischer2015-dc}, NumPy \cite{Van_der_Walt2011-xg}, and SciPy \cite{Virtanen2020-wg}.

\acknowledgments
We are grateful to the Editor Germ\'{a}n Prieto, Piero Poli and the anonymous reviewer for providing helpful comments that improved this study. 
This research was funded by JSPS KAKENHI (Grants 21K21353).

\bibliography{paper_list}

@ARTICLE{Van_der_Walt2011-xg,
  title     = "The {NumPy} array: A structure for efficient numerical
               computation",
  author    = "van der Walt, St{\'e}fan and Colbert, S Chris and Varoquaux,
               Ga{\"e}l",
  journal   = "Comput. Sci. Eng.",
  publisher = "Institute of Electrical and Electronics Engineers (IEEE)",
  volume    =  13,
  number    =  2,
  pages     = "22--30",
  memo     =  mar,
  year      =  2011,
  copyright = "https://ieeexplore.ieee.org/Xplorehelp/downloads/license-information/IEEE.html",
  issn      = "1521-9615, 1558-366X",
  doi       = "10.1109/mcse.2011.37"
}

@ARTICLE{Virtanen2020-wg,
  title    = "{{SciPy} 1.0: fundamental algorithms for scientific computing in
              Python}",
  author   = "Virtanen, Pauli and Gommers, Ralf and Oliphant, Travis E and
              Haberland, Matt and Reddy, Tyler and Cournapeau, David and others",
  journal  = "Nat. Methods",
  volume   =  17,
  number   =  3,
  pages    = "261--272",
  memo     =  mar,
  year     =  2020,
  language = "en",
  issn     = "1548-7091, 1548-7105",
  pmid     = "32015543",
  doi      = "10.1038/s41592-019-0686-2",
  pmc      = "PMC7056644"
}

@ARTICLE{Krischer2015-dc,
  title     = "{ObsPy}: a bridge for seismology into the scientific Python
               ecosystem",
  author    = "Krischer, Lion and Megies, Tobias and Barsch, Robert and
               Beyreuther, Moritz and Lecocq, Thomas and Caudron, Corentin and
               Wassermann, Joachim",
  journal   = "Comput. Sci. Discov.",
  publisher = "IOP Publishing",
  volume    =  8,
  number    =  1,
  pages     = "014003",
  memo      =  may,
  year      =  2015,
  copyright = "http://iopscience.iop.org/page/copyright",
  issn      = "1749-4680, 1749-4699",
  doi       = "10.1088/1749-4699/8/1/014003"
}

@ARTICLE{Ichihara2023-zr,
  title     = "Multiphase turbulent flow explains lightning rings in volcanic
               plumes",
  author    = "Ichihara, Mie and Mininni, Pablo D and Ravichandran, S and
               Cimarelli, Corrado and Vagasky, Chris",
  journal   = "Commun. Earth Environ.",
  publisher = "Springer Science and Business Media LLC",
  volume    =  4,
  number    =  1,
  pages     = "1--12",
  memo     =  nov,
  year      =  2023,
  doi       = "10.1038/s43247-023-01074-z",
  issn      = "2662-4435,2662-4435",
  language  = "en"
}

@ARTICLE{Horiuchi2024-fl,
  title    = "A seismic precursor 15 minutes before the giant eruption of Hunga
              Tonga-Hunga Ha'apai volcano on January 15, 2022",
  author   = "Horiuchi, Takuro and Ichihara, Mie and Nishida, Kiwamu and Kaneko,
              Takayuki",
  journal  = "ESS Open Archive",
  memo    =  jul,
  year     =  2024,
  doi      = "10.22541/essoar.172072911.15953921/v1"
}

@ARTICLE{White2020-oc,
  title    = "{PyKonal}: A Python Package for Solving the Eikonal Equation in
              Spherical and Cartesian Coordinates Using the Fast Marching
              Method",
  author   = "White, Malcolm C A and Fang, Hongjian and Nakata, Nori and
              Ben-Zion, Yehuda",
  journal  = "Seismol. Res. Lett.",
  memo    =  jun,
  year     =  2020,
  issn     = "0895-0695",
  doi      = "10.1785/0220190318"
}

@ARTICLE{Tromp1992-qt,
  title     = "Variational principles for surface wave propagation on a
               laterally heterogeneous {Earth---II}. Frequency-domain {JWKB}
               theory",
  author    = "Tromp, Jeroen and Dahlen, F A",
  journal   = "Geophys. J. Int.",
  publisher = "Oxford Academic",
  volume    =  109,
  number    =  3,
  pages     = "599--619",
  memo     =  jun,
  year      =  1992,
  language  = "en",
  issn      = "0956-540X",
  doi       = "10.1111/j.1365-246X.1992.tb00120.x"
}

@BOOK{Pedlosky2003-va,
  title     = "Waves in the Ocean and Atmosphere: Introduction to Wave Dynamics",
  author    = "Pedlosky, Joseph",
  publisher = "Springer Science \& Business Media",
  memo     =  jun,
  year      =  2003,
  language  = "en",
  isbn      = "9783540003403"
}

@ARTICLE{Diaz2023-cc,
  title     = "Seismic record of a long duration dispersive signal after the 15
               January 2022 {Hunga-Tonga} eruption",
  author    = "Diaz, Jordi",
  journal   = "Seismica",
  publisher = "McGill University Library and Archives",
  volume    =  2,
  number    =  2,
  memo     =  nov,
  year      =  2023,
  issn      = "2816-9387, 2816-9387",
  doi       = "10.26443/seismica.v2i2.1033"
}

@BOOK{Dahlen1998-hr,
  title     = "{Theoretical Global Seismology}",
  author    = "Dahlen, F A and Tromp, J",
  publisher = "Princeton University Press",
  pages     = "1025pp",
  year      =  1998,
  address   = "Princeton"
}

@ARTICLE{Nishida2019-eb,
  title    = "Seismic Observation of Tsunami at Island Broadband Stations",
  author   = "Nishida, Kiwamu and Maeda, Takuto and Fukao, Yoshio",
 
  journal  = "J. Geophys. Res.",
  volume   =  124,
  number   =  2,
  pages    = "1910--1928",
  memo    =  feb,
  year     =  2019,
  issn     = "2169-9313",
  doi      = "10.1029/2018JB016833"
}

@ARTICLE{Tonegawa2018-zp,
  title    = "Excitation Location and Seasonal Variation of Transoceanic
              Infragravity Waves Observed at an Absolute Pressure Gauge Array",
  author   = "Tonegawa, T and Fukao, Y and Shiobara, H and Sugioka, H and Ito,
              A and Yamashita, M",
  journal  = "J. Geophys. Res. C: Oceans",
  pages    = "40--52",
  year     =  2018,
  issn     = "2169-9275",
  doi      = "10.1002/2017JC013488"
}

@ARTICLE{Yuan2005-ir,
  title    = "Seismic monitoring of the Indian Ocean tsunami",
  author   = "Yuan, X and Kind, R and Pedersen, H A",
  journal  = "Geophys. Res. Lett.",
  volume   =  32,
  number   =  15,
  pages    = "L15308",
  year     =  2005,
  issn     = "0094-8276",
  doi      = "10.1029/2005GL023464"
}

@ARTICLE{Hanson2005-cz,
  title    = "Dispersive and reflected tsunami signals from the 2004 Indian
              Ocean tsunami observed on hydrophones and seismic stations",
  author   = "Hanson, Jeffrey A and Bowman, J Roger",
  journal  = "Geophys. Res. Lett.",
  volume   =  32,
  number   =  17,
  pages    = "1--5",
  year     =  2005,
  issn     = "0094-8276",
  doi      = "10.1029/2005GL023783"
}

@ARTICLE{Longuet-Higgins1967-kr,
  title     = "On the trapping of wave energy round islands",
  author    = "Longuet-Higgins, M S",
  journal   = "J. Fluid Mech.",
  publisher = "University of Tokyo Libraries",
  volume    =  29,
  number    =  04,
  pages     = "781--821",
  year      =  1967,
  issn      = "0022-1120",
  doi       = "10.1017/S0022112067001181"
}

@ARTICLE{Okal2007-vn,
  title    = "Seismic Records of the 2004 Sumatra and Other Tsunamis: A
              Quantitative Study",
  author   = "Okal, Emile A",
  journal  = "Pure Appl. Geophys.",
  volume   =  164,
  number   =  2,
  pages    = "325--353",
  memo    =  mar,
  year     =  2007,
  issn     = "0033-4553, 1420-9136",
  doi      = "10.1007/s00024-006-0181-4"
}

@ARTICLE{Okal2007-sc,
  title    = "Quantification of hydrophone records of the 2004 Sumatra Tsunami",
  author   = "Okal, Emile A and Talandier, Jacques and Reymond, Dominique",
  journal  = "Pure Appl. Geophys.",
  volume   =  164,
  number   = "2-3",
  pages    = "309--323",
  year     =  2007,
  issn     = "0033-4553",
  doi      = "10.1007/s00024-006-0165-4"
}

@ARTICLE{Seabrook2023-nl,
  title    = "Volcaniclastic density currents explain widespread and diverse
              seafloor impacts of the 2022 Hunga Volcano eruption",
  author   = "Seabrook, Sarah and Mackay, Kevin and Watson, Sally J and Clare,
              Michael A and Hunt, James E and Yeo, Isobel A and Lane, Emily M
              and Clark, Malcolm R and Wysoczanski, Richard and Rowden, Ashley
              A and Kula, Taaniela and Hoffmann, Linn J and Armstrong, Evelyn
              and Williams, Michael J M",
  journal  = "Nat. Commun.",
  volume   =  14,
  number   =  1,
  pages    = "7881",
  memo    =  nov,
  year     =  2023,
  language = "en",
  issn     = "2041-1723",
  pmid     = "38036504",
  doi      = "10.1038/s41467-023-43607-2",
  pmc      = "PMC10689732"
}

@ARTICLE{Mizutani2023-gb,
  title     = "Source estimation of the tsunami later phases associated with
               the 2022 Hunga Tonga volcanic eruption",
  author    = "Mizutani, Ayumu and Yomogida, Kiyoshi",
  journal   = "Geophys. J. Int.",
  publisher = "Oxford Academic",
  volume    =  234,
  number    =  3,
  pages     = "1885--1902",
 memo    =  apr,
  year      =  2023,
  issn      = "0956-540X",
  doi       = "10.1093/gji/ggad174"
}

@ARTICLE{Ravanelli2023-xf,
  title     = "Ocean‐ionosphere disturbances due to the 15 {January} 2022 {Hunga‐Tonga} {Hunga‐Ha'apai} eruption",
  author    = "Ravanelli, M and Astafyeva, E and Munaibari, E and Rolland, L and Mikesell, T D",
  journal   = "Geophys. Res. Lett.",
  publisher = "American Geophysical Union (AGU)",
  volume    =  50,
  number    =  10,
 memo    =  may,
  year      =  2023,
  copyright = "http://creativecommons.org/licenses/by-nc-nd/4.0/",
  language  = "en",
  issn      = "0094-8276, 1944-8007",
  doi       = "10.1029/2022gl101465"
}

@ARTICLE{Omira2022-al,
  title     = "Global Tonga tsunami explained by a fast-moving atmospheric
               source",
  author    = "Omira, R and Ramalho, R S and Kim, J and Gonz{\'a}lez, P J and
               Kadri, U and Miranda, J M and others",
  journal   = "Nature",
  pages     = "1--7",
 memo    =  jun,
  year      =  2022,
  language  = "en",
  issn      = "0028-0836",
  doi       = "10.1038/s41586-022-04926-4"
}

@MISC{Amante2009-jn,
  title     = "{ETOPO1} Global Relief Model converted to {PanMap} layer format",
  author    = "Amante, C and Eakins, B W",
  publisher = "PANGAEA",
  memo     =  sep,
  year      =  2009,
}

@ARTICLE{Matoza2022-tt,
  title    = "Atmospheric waves and global seismoacoustic observations of the
              January 2022 Hunga eruption, Tonga",
  author   = "Matoza, Robin S and Fee, David and Assink, Jelle D and Iezzi,
              Alexandra M and Green, David N and Kim, Keehoon and others",
  journal  = "Science",
  volume   =  377,
  number   =  6601,
  pages    = "95--100",
  memo    =  jul,
  year     =  2022,
  language = "en",
  eprint   = "https://www.science.org/doi/pdf/10.1126/science.abo7063",
  issn     = "0036-8075, 1095-9203",
  pmid     = "35549311",
  doi      = "10.1126/science.abo7063"
}

@ARTICLE{Kubota2022-fb,
  title    = "Global fast-traveling tsunamis driven by atmospheric {Lamb} waves on the 2022 {Tonga} eruption",
  author   = "Kubota, Tatsuya and Saito, Tatsuhiko and Nishida, Kiwamu",
  journal  = "Science",
  volume   =  377,
  number   =  6601,
  pages    = "91--94",
  memo    =  jul,
  year     =  2022,
  language = "en",
  eprint   = "https://www.science.org/doi/pdf/10.1126/science.abo4364",
  issn     = "0036-8075, 1095-9203",
  pmid     = "35549307",
  doi      = "10.1126/science.abo4364"
}

@ARTICLE{Astafyeva2022-xy,
  title     = "The 15 {January} 2022 {Hunga Tonga} eruption history as inferred from ionospheric observations",
  author    = "Astafyeva, E and Maletckii, B and Mikesell, T D and Munaibari, E
               and Ravanelli, M and Coisson, P and others", 
  journal   = "Geophys. Res. Lett.",
  publisher = "American Geophysical Union (AGU)",
  memo     =  may,
  year      =  2022,
  copyright = "http://onlinelibrary.wiley.com/termsAndConditions\#vor",
  language  = "en",
  issn      = "0094-8276, 1944-8007",
  doi       = "10.1029/2022gl098827"
}

@ARTICLE{Vergoz2022-ch,
  title    = "{IMS} observations of infrasound and acoustic-gravity waves
              produced by the {January} 2022 volcanic eruption of {Hunga, Tonga:} {A
              global analysis}",
  author   = "Vergoz, J and Hupe, P and Listowski, C and Le Pichon, A and
              Garc{\'e}s, M A and Marchetti, E and others",
  journal  = "Earth Planet. Sci. Lett.",
  volume   =  591,
  pages    = "117639",
  memo    =  aug,
  year     =  2022,
  issn     = "0012-821X",
  doi      = "10.1016/j.epsl.2022.117639"
}

@ARTICLE{Wright2022-lh,
  title    = "{Surface-to-space atmospheric waves from Hunga {Tonga-Hunga}
              Ha'apai eruption}",
  author   = "Wright, Corwin J and Hindley, Neil P and Alexander, M Joan and
              Barlow, Mathew and Hoffmann, Lars and Mitchell, Cathryn N and
              others",
  journal  = "Nature",
  memo    =  jun,
  year     =  2022,
  language = "en",
  issn     = "0028-0836, 1476-4687",
  pmid     = "35772670",
  doi      = "10.1038/s41586-022-05012-5"
}

@ARTICLE{Thurin2023-gw,
  title    = "Comparison of force and moment tensor estimations of subevents
              during the 2022 {Hunga-Tonga} submarine volcanic eruption",
  author   = "Thurin, J and Tape, C",
  journal  = "Geophys. J. Int.",
  volume   =  235,
  number   =  2,
  pages    = "1959--1981",
  memo    =  nov,
  year     =  2023,
  doi      = "10.1093/gji/ggad323"
}

@ARTICLE{Brodsky1999-hw,
  title     = "A seismically constrained mass discharge rate for the initiation of the {May} 18, 1980 {Mount St. Helens} eruption",
  author    = "Brodsky, Emily E and Kanamori, Hiroo and Sturtevant, Bradford",
  journal   = "J. Geophys. Res.",
  publisher = "American Geophysical Union (AGU)",
  volume    =  104,
  number    = "B12",
  pages     = "29387--29400",
  memo     =  dec,
  year      =  1999,
  language  = "en",
  issn      = "0148-0227, 2156-2202",
  doi       = "10.1029/1999jb900308"
}

@ARTICLE{Poli2022-ef,
  title     = "Rapid characterization of large volcanic eruptions: Measuring
               the impulse of the Hunga Tonga ha'apai explosion from
               teleseismic waves",
  author    = "Poli, Piero and Shapiro, Nikolai M",
  journal   = "Geophys. Res. Lett.",
  publisher = "American Geophysical Union (AGU)",
  volume    =  49,
  number    =  8,
  memo     =  apr,
  year      =  2022,
  copyright = "http://onlinelibrary.wiley.com/termsAndConditions\#vor",
  language  = "en",
  issn      = "0094-8276, 1944-8007",
  doi       = "10.1029/2022gl098123"
}

@ARTICLE{Kubo2022-le,
  title     = "Ocean-wave phenomenon around Japan due to the 2022 {Tonga} eruption observed by the wide and dense ocean-bottom pressure gauge networks",
  author    = "Kubo, Hisahiko and Kubota, Tatsuya and Suzuki, Wataru and Aoi, Shin and Sandanbata, Osamu and Chikasada, Naotaka and Ueda, Hideki",
  journal   = "Earth Planets Space",
  publisher = "SpringerOpen",
  volume    =  74,
  number    =  1,
  pages     = "1--11",
  memo     =  jul,
  year      =  2022,
  language  = "en",
  issn      = "1343-8832",
  doi       = "10.1186/s40623-022-01663-w"
}

@ARTICLE{Nishikawa2022-ep,
  title    = "Observation and simulation of atmospheric gravity waves exciting subsequent tsunami along the coastline of Japan after {Tonga} explosion event",
  author   = "Nishikawa, Yasuhiro and Yamamoto, Masa-Yuki and Nakajima, Kensuke
              and Hamama, Islam and Saito, Hiroaki and Kakinami, Yoshihiro and
              others",
  journal  = "Sci. Rep.",
  volume   =  12,
  number   =  1,
  pages    = "22354",
  memo    =  dec,
  year     =  2022,
  language = "en",
  issn     = "2045-2322",
  pmid     = "36572667",
  doi      = "10.1038/s41598-022-25854-3",
  pmc      = "PMC9792542"
}

@ARTICLE{Millan2022-xl,
  title    = "The {Hunga} {Tonga-Hunga} {Ha'apai} Hydration of the Stratosphere",
  author   = "Mill{\'a}n, L and Santee, M L and Lambert, A and Livesey, N J and
              Werner, F and Schwartz, M J and others",
  journal  = "Geophys. Res. Lett.",
  volume   =  49,
  number   =  13,
  pages    = "e2022GL099381",
  memo    =  jul,
  year     =  2022,
  language = "en",
  issn     = "0094-8276",
  pmid     = "35865735",
  doi      = "10.1029/2022GL099381",
  pmc      = "PMC9285945"
}

@ARTICLE{Borrero2022-gk,
  title    = "{Tsunami Runup and Inundation in Tonga from the January 2022
              eruption of Hunga Volcano}",
  author   = "Borrero, Jose C and Cronin, Shane J and Latu'ila, Folauhola
              Helina and Tukuafu, Pupunu and Heni, Nikolasi and Tupou, Ana Maea
              and others",
  memo    =  sep,
  year     =  2022,
  language = "en",
  doi      = "10.21203/rs.3.rs-2044907/v1"
}

@ARTICLE{Le_Bras2023-gs,
  title     = "The {Hunga} {Tonga--Hunga} {Ha'apai} {Eruption} of 15 {January} 2022:
               Observations on the {International Monitoring System} ({IMS})
               {Hydroacoustic Stations and Synergy with Seismic and Infrasound Sensors}",
  author    = "Le Bras, Ronan J and Zampolli, Mario and Metz, Dirk and
               Haralabus, Georgios and Bittner, Paulina and Villarroel, Marcela
               and Matsumoto, Hiroyuki and Graham, Gerhard and {\"O}zel, Nurcan
               Meral",
  journal   = "Seismol. Res. Lett.",
  publisher = "GeoScienceWorld",
  volume    =  94,
  number    = "2A",
  pages     = "578--588",
  memo     =  mar,
  year      =  2023,
  issn      = "0895-0695",
  doi       = "10.1785/0220220240"
}

@ARTICLE{Schoeberl2022-tj,
  title     = "Analysis and impact of the Hunga {Tonga‐Hunga} {Ha'apai} stratospheric water vapor plume",
  author    = "Schoeberl, M R and Wang, Y and Ueyama, R and Taha, G and Jensen, E and Yu, W",
  journal   = "Geophys. Res. Lett.",
  publisher = "American Geophysical Union (AGU)",
  volume    =  49,
  number    =  20,
  memo     =  oct,
  year      =  2022,
  copyright = "http://creativecommons.org/licenses/by/4.0/",
  language  = "en",
  issn      = "0094-8276, 1944-8007",
  doi       = "10.1029/2022gl100248"
}

@ARTICLE{Garza-Giron2023-bh,
  title    = "Solid Earth-atmosphere interaction forces during the 15 {January} 2022 {Tonga} eruption",
  author   = "Garza-Gir{\'o}n, Ricardo and Lay, Thorne and Pollitz, Frederick
              and Kanamori, Hiroo and Rivera, Luis",
  journal  = "Sci Adv",
  volume   =  9,
  number   =  2,
  pages    = "eadd4931",
  memo    =  jan,
  year     =  2023,
  language = "en",
  issn     = "2375-2548",
  pmid     = "36630503",
  doi      = "10.1126/sciadv.add4931"
}

@ARTICLE{Watada2023-eb,
  title     = "Detection of air temperature and wind changes synchronized with
               the {Lamb} wave from the 2022 {Tonga} volcanic eruption",
  author    = "Watada, Shingo and Imanishi, Yuichi and Tanaka, Kenji",
  journal   = "Geophys. Res. Lett.",
  publisher = "American Geophysical Union (AGU)",
  memo     =  jan,
  year      =  2023,
  copyright = "http://creativecommons.org/licenses/by/4.0/",
  language  = "en",
  issn      = "0094-8276, 1944-8007",
  doi       = "10.1029/2022gl100884"
}

@ARTICLE{Watanabe2022-bu,
  title     = "{First Detection of the Pekeris Internal Global Atmospheric Resonance: Evidence from the 2022 {Tonga} Eruption and from Global Reanalysis Data}",
  author    = "Watanabe, Shingo and Hamilton, Kevin and Sakazaki, Takatoshi and Nakano, Masuo",
  journal   = "J. Atmos. Sci.",
  publisher = "American Meteorological Society",
  volume    =  79,
  number    =  11,
  pages     = "3027--3043",
  memo     =  nov,
  year      =  2022,
  language  = "en",
  issn      = "0022-4928, 1520-0469",
  doi       = "10.1175/JAS-D-22-0078.1"
}

@ARTICLE{Webb2010-vb,
  title     = "{Shallow-Water} Broadband {OBS} Seismology",
  author    = "Webb, Spahr C and Crawford, Wayne C",
  journal   = "Bull. Seismol. Soc. Am.",
  publisher = "GeoScienceWorld",
  volume    =  100,
  number    =  4,
  pages     = "1770--1778",
  memo     =  aug,
  year      =  2010,
  issn      = "0037-1106",
  doi       = "10.1785/0120090203"
}

@ARTICLE{Janiszewski2022-zp,
  title     = "Broad-band ocean bottom seismometer noise properties",
  author    = "Janiszewski, Helen A and Eilon, Z and Russell, J B and Brunsvik,
               B and Gaherty, J B and Mosher, S G and others",
  journal   = "Geophys. J. Int.",
  publisher = "Oxford Academic",
  volume    =  233,
  number    =  1,
  pages     = "297--315",
  memo     =  nov,
  year      =  2022,
  issn      = "0956-540X",
  doi       = "10.1093/gji/ggac450"
}

@ARTICLE{Yuen2022-na,
  title     = "Under the surface: Pressure-induced planetary-scale waves,
               volcanic lightning, and gaseous clouds caused by the submarine
               eruption of Hunga {Tonga-Hunga} Ha'apai volcano",
  author    = "Yuen, David A and Scruggs, Melissa A and Spera, Frank J and
               Zheng, Yingcai and Hu, Hao and McNutt, Stephen R and Thompson,
               Glenn and Mandli, Kyle and Keller, Barry R and Wei, Songqiao
               Shawn and Peng, Zhigang and Zhou, Zili and Mulargia, Francesco
               and Tanioka, Yuichiro",
  journal   = "Earthquake Research Advances",
  publisher = "Elsevier BV",
  volume    =  2,
  number    =  3,
  pages     = "100134",
  memo     =  jul,
  year      =  2022,
  copyright = "http://creativecommons.org/licenses/by/4.0/",
  language  = "en",
  issn      = "2772-4670",
  doi       = "10.1016/j.eqrea.2022.100134"
}

@ARTICLE{Ringler2022-ti,
  title     = "The global seismographic network reveals atmospherically coupled normal modes excited by the 2022 {Hunga Tonga} eruption",
  author    = "Ringler, A T and Anthony, R E and Aster, R C and Taira, T and
               Shiro, B R and Wilson, D C and others",
  journal   = "Geophys. J. Int.",
  publisher = "Oxford Academic",
  volume    =  232,
  number    =  3,
  pages     = "2160--2174",
  memo     =  nov,
  year      =  2022,
  issn      = "0956-540X",
  doi       = "10.1093/gji/ggac284"
}

@ARTICLE{Tonegawa2023-gm,
  title    = "Mesospheric pressure source from the 2022 {Hunga, Tonga} eruption excites 3.6-{mHz} air-sea coupled waves",
  author   = "Tonegawa, Takashi and Fukao, Yoshio",
  journal  = "Sci Adv",
  volume   =  9,
  number   =  26,
  pages    = "eadg8036",
  memo    =  jun,
  year     =  2023,
  language = "en",
  issn     = "2375-2548",
  pmid     = "37379387",
  doi      = "10.1126/sciadv.adg8036",
  pmc      = "PMC10306286"
}

@ARTICLE{Purkis2023-ec,
  title    = "The 2022 {Hunga-Tonga} megatsunami: Near-field simulation of a
              once-in-a-century event",
  author   = "Purkis, Sam J and Ward, Steven N and Fitzpatrick, Nathan M and
              Garvin, James B and Slayback, Dan and Cronin, Shane J and
              others",
  journal  = "Sci Adv",
  volume   =  9,
  number   =  15,
  pages    = "eadf5493",
  memo    =  apr,
  year     =  2023,
  language = "en",
  issn     = "2375-2548",
  pmid     = "37058570",
  doi      = "10.1126/sciadv.adf5493"
}

\end{document}